\documentclass{article}[12pt]

\usepackage{graphicx}

\def\zehn{10 }\def\acht{8 }\def\sieben{7 }\def\fuenf{5 }\def\zwoelf{10 }
\def\zweiundzwanzig{10 }\def\siebzehn{10 }
\def\makeatletter{\catcode`\@=11\relax}

\makeatletter
\ifcase\@ptsize
   \def\zehnpt{9.8}	
   \def\smallpt{9.2}	\def\footnotesizept{8}	\def\scriptsizept{7}
   \def\tinypt{5}	\def\largept{12}	\def\Largept{14.5}
   \def\LARGEpt{17.5}	\def\hugept{20.5}	\def\Hugept{25}
\or
   \def\zehnpt{11.01}	
   \def\smallpt{9.8}	\def\footnotesizept{9.2}\def\scriptsizept{8}
   \def\tinypt{6}	\def\largept{12}	\def\Largept{14.5}
   \def\LARGEpt{17.5}	\def\hugept{20.5}	\def\Hugept{25}
\or
   \def\zehnpt{12}	
   \def\smallpt{11.01}	\def\footnotesizept{9.8}	\def\scriptsizept{8}
   \def\tinypt{6}	\def\largept{14.5}	\def\Largept{17.5}
   \def\LARGEpt{20.5}	\def\hugept{25.01}		\def\Hugept{25.01}
\fi
\def\XNEUERSATZ#1#2#3#4{%
\global\expandafter\font\csname #1#4\endcsname=\FONTSATZ#2 at #3 pt\csname #1#4\endcsname}%
\def\YNEUERSATZ#1#2#3#4{%
\global\expandafter\font\csname #1#4\endcsname=\FONTSATZ#2 scaled #3\csname #1#4\endcsname}%
\def\NEUERSATZ#1#2{%
\def\LEER{}\ifx\SKALIERUNG\LEER{}\XNEUERSATZ{\S@G}{#1}{#2}{\ZEICHENSATZ}%
\else%
\dimendef\SCALA=230\SCALA=#2 pt\divide\SCALA by 128\multiply\SCALA by \SKALIERUNG%
\countdef\scala=230\scala=\SCALA\divide\scala by 5120\edef\SCALA{\the\scala}%
\YNEUERSATZ{\S@G}{ }{\SCALA}{\ZEICHENSATZ}%
\fi}

\def\setze#1#2#3#4{%
\gdef\ZEICHENSATZ{#1}\gdef\FONTSATZ{#2}\gdef\SKALIERUNG{#3}%
#4%
\global\expandafter\newfam\csname X#1\endcsname%
{\@small\NEUERSATZ{\zehn}{\smallpt}}
{\@footnote\NEUERSATZ{\acht}{\footnotesizept}}
{\@large\NEUERSATZ{\zwoelf}{\largept}}
{\@Large\NEUERSATZ{\siebzehn}{\Largept}}
{\@LARGE\NEUERSATZ{\siebzehn}{\LARGEpt}}
{\@huge\NEUERSATZ{\zweiundzwanzig}{\hugept}}
{\@Huge\NEUERSATZ{\zweiundzwanzig}{\Hugept}}
\@scriptsize\NEUERSATZ{\sieben}{\scriptsizept}%
\global\expandafter\scriptfont\csname X#1\endcsname=\expandafter\csname \S@G#1\endcsname%
\@tiny\NEUERSATZ{\fuenf}{\tinypt}%
\global\expandafter\scriptscriptfont\csname X#1\endcsname=\expandafter\csname \S@G#1\endcsname%
\@zehn\NEUERSATZ{\zehn}{\zehnpt}
\global\expandafter\textfont\csname X#1\endcsname=\expandafter\csname \S@G#1\endcsname%
\global\expandafter\def\csname #1\endcsname{\edef\ZEICHENSATZ{#1}\expandafter\fam\csname X#1\endcsname%
\expandafter\csname \S@G\ZEICHENSATZ\endcsname}%
}

\def\@zehn{\def\S@G{zehn}}
\def\@small{\def\S@G{small}}
\def\@footnote{\def\S@G{footnote}}
\def\@scriptsize{\def\S@G{scriptsize}}
\def\@tiny{\def\S@G{tiny}}
\def\@large{\def\S@G{large}}
\def\@Large{\def\S@G{Large}}
\def\@LARGE{\def\S@G{LARGE}}
\def\@huge{\def\S@G{huge}}
\def\@Huge{\def\S@G{Huge}}
\def\Xnormalsize{\@zehn\expandafter\csname \S@G\ZEICHENSATZ\endcsname}
\def\Xsmall{\@small\expandafter\csname \S@G\ZEICHENSATZ\endcsname}
\def\Xfootnotesize{\@footnote\expandafter\csname \S@G\ZEICHENSATZ\endcsname}
\def\Xscriptsize{\@scriptsize\expandafter\csname \S@G\ZEICHENSATZ\endcsname}
\def\Xtiny{\@tiny\expandafter\csname \S@G\ZEICHENSATZ\endcsname}
\def\Xlarge{\@large\expandafter\csname \S@G\ZEICHENSATZ\endcsname}
\def\XLarge{\@Large\expandafter\csname \S@G\ZEICHENSATZ\endcsname}
\def\XLARGE{\@LARGE\expandafter\csname \S@G\ZEICHENSATZ\endcsname}
\def\Xhuge{\@huge\expandafter\csname \S@G\ZEICHENSATZ\endcsname}
\def\XHuge{\@Huge\expandafter\csname \S@G\ZEICHENSATZ\endcsname}

\setze{doppelt}{msbm}{}{}
\setze{fmath}{cmmib}{}{\def\zehn{9 }\def\siebzehn{9 }\def\zweiundzwanzig{9 }}

\chardef\FC=82
\newcommand{\real}{\hbox{\doppelt\FC}}
\chardef\ED=67

\chardef\FK=90

\advance\voffset by -0.25cm

\newcommand{\natureline}{\begin{tabular}{l}
                         \hspace{5.85in} \\
                         \hline \\
                         \end{tabular}}
\newcommand{\und}{$\&$ }

\begin{document}

\begin{center}
{\large \bf Stochastic models which separate fractal\\[0.22em] 
            dimension and Hurst effect} 

\medskip
\noindent
{Tilmann Gneiting$^1$ and Martin Schlather$^2$}

\medskip
\noindent
$^1${\em{Department} of Statistics, University of Washington,  
    Seattle, Washington 98195, USA} \\
$^2${\em{Soil} Physics Group, Universit\"at Bayreuth, 95440 Bayreuth, 
    Germany} 
\end{center}

\begin{abstract}
Fractal behavior and long-range dependence have been observed in an
astonishing number of physical systems. Either phenomenon has been
modeled by self-similar random functions, thereby implying a linear
relationship between fractal dimension, a measure of roughness, and
Hurst coefficient, a measure of long-memory dependence. This letter
introduces simple stochastic models which allow for any combination of
fractal dimension and Hurst exponent. We synthesize images from these
models, with arbitrary fractal properties and power-law correlations,
and propose a test for self-similarity.

\medskip
\noindent
PACS numbers: 02.50.Ey, 02.70-c, 05.40-a, 05.45.Df
\end{abstract} 

\medskip
{\em I. Introduction.} Following Mandelbrot's seminal essay [1],
fractal-based analyses of time series, profiles, and natural or
man-made surfaces have found extensive applications in almost all
scientific disciplines [2--5]. The fractal dimension, $D$, of a
profile or surface is a measure of roughness, with $D \in [n,n+1)$ for
a surface in $n$-dimensional space and higher values indicating
rougher surfaces. Long-memory dependence or persistence in time series
[6--8] or spatial data [9--11] is associated with power-law
correlations and often referred to as Hurst effect. Scientists in
diverse fields observed empirically that correlations between
observations that are far apart in time or space decay much slower
than would be expected from classical stochastic models. Long-memory
dependence is characterized by the Hurst coefficient, $H$. In
principle, fractal dimension and Hurst coefficient are independent of
each other: fractal dimension is a local property, and long-memory
dependence is a global characteristic. Nevertheless, the two notions
are closely linked in much of the scientific literature. This stems
from the success of self-similar models such as fractional Gaussian
noise and fractional Brownian motion [12] in modeling and explaining
either phenomenon. For self-similar processes, the local properties
are reflected in the global ones, resulting in the celebrated
relationship
\begin{equation} \label{eq:DandH}
D = n + 1 - H 
\end{equation}
between fractal dimension, $D$, and Hurst coefficient, $H$, for a
self-similar surface in $n$-dimensional space [1,3]. Long-memory
dependence, or persistence, is associated with the case $H \in
(\frac{1}{2},1)$ and therefore linked to surfaces with low fractal
dimensions. Rougher surfaces with higher fractal dimensions occur for
antipersistent processes with $H \in (0,\frac{1}{2})$. Self-similarity
is undoubtedly a natural assumption for many physical, geological, and
biological systems. Owing to its intuitive appeal and a lack of
suitable alternatives, self-similarity and the linear relationship
(\ref{eq:DandH}) are believed to be warranted by a large number of
real-world data sets.

The stochastic models presented here provide a fresh perspective,
since they allow for any combination of fractal dimension, $D$, and
Hurst exponent, $H$. The models are very simple, have only two
exponents, and allow for the straightforward synthesis of images with
arbitrary fractal properties and power-law correlations. We call for a
critical assessment of self-similar models and of the relationship
(\ref{eq:DandH}) through joint measurements of $D$ and $H$ in physical
systems.

\bigskip
{\em II. Stationary processes.} This section recalls some basic facts
for reference below. In the interest of a clear presentation, we
restrict ourselves to a discussion of stationary, standard Gaussian
[13] random functions $Z(x)$, $x \in \real^n$, which are characterized
by their correlation function,
\begin{equation}
c(h) = \; < \! Z(x) Z(x+h) \! >, \qquad\quad h \in \real^n.
\end{equation}
The behavior of the correlation function at $h=0$ determines the local
properties of the realizations. Specifically, if
\begin{equation} \label{eq:alpha} 
1-c(h) \sim |h|^\alpha \quad \mbox{as} \quad h \to 0 
\end{equation} 
for some $\alpha \in (0,2]$, then the realizations of the random function 
have fractal dimension
\begin{equation} \label{eq:D}
D = n + 1 - \frac{\alpha}{2}
\end{equation}
with probability one [14]. Similarly, the asymptotic behavior of the
correlation function at infinity determines the presence or absence of
long-range dependence. Long-memory processes are associated with
power-law correlations,
\begin{equation} \label{eq:beta} 
c(h) \sim |h|^{-\beta} \quad \mbox{as} \quad |h| \to \infty,
\end{equation} 
and if $\beta \in (0,2)$, the behavior is frequently expressed in
terms of the Hurst coefficient,
\begin{equation} \label{eq:H} 
H = 1 - \frac{\beta}{2}. 
\end{equation} 
The asymptotic relationships (\ref{eq:alpha}) and (\ref{eq:beta}) can
be expressed equivalently in terms of the spectral density and its
behavior at infinity and zero, respectively. The traditional
stationary, self-similar stochastic process is fractional Gaussian
noise [12], that is, the Gaussian process with correlation function
\begin{equation} \label{eq:fGn} 
c(h) = \frac{1}{2} \left( |h+1|^{2H} - 2 |h|^{2H} +|h-1|^{2H} \right) ,  
       \qquad\quad h \in \real,
\end{equation} 
where $H \in (0,1)$ is the Hurst coefficient. Then $1-c(h) \sim |h|^{2H}$
as $h \to 0$ and
\begin{equation} \label{eq:fGn_as} 
\frac{c(h)}{H (2H-1) |h|^{-(2-2H)}} \to 1
   \quad \mbox{as} \quad |h| \to \infty;
\end{equation} 
hence, the linear relationship (\ref{eq:DandH}) holds with $n=1$. The
case $H \in (\frac{1}{2},1)$ is associated with positive correlations,
persistent processes, and low fractal dimensions; if $H \in
(0,\frac{1}{2})$ we find negative correlations, antipersistent
processes, and high fractal dimensions. In other words, the assumption
of statistical self-similarity determines the relationships between
local and global behavior, or fractal dimension and Hurst effect. By
way of contrast, the stochastic models presented hereinafter allow for
any combination of fractal dimension and Hurst coefficient.

\bigskip
{\em III. Cauchy class.} The Cauchy class consists of the stationary
Gaussian random processes $Z(x)$, $x \in \real^n$, with correlation
function
\begin{equation} \label{eq:Cauchy}
c(h) = \left( 1 + |h|^\alpha \right)^{-\beta/\alpha}, 
  \qquad\quad h \in \real^n,
\end{equation}
for any combination of the parameters $\alpha \in (0,2]$ and $\beta
\geq 0$. It provides flexible power-law correlations and generalizes
stochastic models recently discussed and synthesized in geostatistics
[15], physics [11,16], hydrology [17], and time series analysis
[18--19]. These works consider time series (in discrete time) only, or
they restrict $\alpha$ to 1 or 2. The special case $\alpha=2$ has been
known as Cauchy model [15], and we refer to the general case, $\alpha
\in (0,2]$, as Cauchy class. The correlation function
(\ref{eq:Cauchy}) behaves like (\ref{eq:alpha}) and (\ref{eq:beta}) as
$h \to 0$ and $|h| \to \infty$, respectively. Thus, the realizations
of the associated random process have fractal dimension $D$, as given
by (\ref{eq:D}); and if $\beta \in (0,2)$ the Hurst coefficient, $H$,
is given by (\ref{eq:H}). In particular, $D$ and $H$ may vary
independently.  Figure 1 illustrates two-dimensional realizations of
the Cauchy class for various values of $\alpha$ and $\beta$. In each
row, $\beta = 2-2H$ is constant; but from left to right $\alpha$
increases, that is, the fractal dimension, $D = 3 - \frac{\alpha}{2}$,
decreases. In each column, the fractal dimension is constant, but from
top to bottom the Hurst coefficient decreases. These values of $D$ and
$H$ are the theoretical quantities as determined by the correlation
function. The measured values for the realizations differ from the
theoretical ones, due to chance variability and the discrete nature of
simulations, but only slightly so. We used the turning bands method
with line simulations by the circulant embedding approach [20] to
generate the realizations. The code is publicly available [21] and
allows, for the first time, for the straightforward synthesis of
images with any given combination of fractal dimension and Hurst
coefficient.

\bigskip
{\em IV. Modified Cauchy class.} The Cauchy class allows for any
positive parameter $\beta$ in the power-law (\ref{eq:beta}), and the
correlations are always positive. In contrast, fractional Gaussian
noise can only model power-laws with $\beta \in (0,2)$, and the
correlations eventually become negative if $H \in (0,\frac{1}{2})$ or
$\beta \in (1,2)$. We consider the positive correlations to be an
advantage of the Cauchy model, since positive power-laws are
ubiquitous in the physical, geological, and biological
sciences. Nevertheless, we present another stochastic model, the
modified Cauchy class, which allows for any combination of fractal
dimension and Hurst coefficient and also features the transition from
persistence to antipersistence. The modified Cauchy class consists of
the stationary Gaussian random processes $Z(x)$, $x \in \real$, with
correlation function
\begin{equation} \label{eq:mod_Cauchy}
c(h) = \left( 1 + |h|^\alpha \right)^{-(\beta/\alpha)-1} \left( 1 +
       (1-\beta) |h|^\alpha \right), \qquad\quad h \in \real,
\end{equation}
where $\alpha \in (0,2]$ and $\beta \geq 0$ [22]. In the same way as
(\ref{eq:Cauchy}), the correlation function (\ref{eq:mod_Cauchy})
behaves like (\ref{eq:alpha}) and (\ref{eq:beta}) as $h \to 0$ and
$|h| \to \infty$, respectively, yielding the same conclusions for the
fractal dimension, $D$, given by (\ref{eq:D}), and Hurst coefficient,
$H$, given by (\ref{eq:H}). Furthermore, there is a transition from
positive to negative correlations, or persistence to antipersistence,
respectively, depending on whether $\beta$ is smaller or greater than
1. Similarly to fractional Gaussian noise, (\ref{eq:mod_Cauchy}) is a
valid correlation function in $\real$, but not in the general
Euclidean space $\real^n$ ($n>1$). Figure 2 illustrates realizations
of the modified Cauchy class. The graphs along the subdiagonal
correspond to parameter combinations with $\alpha+\beta=2$, or
$D=2-H$, the same relationship as for self-similar processes. The
graphs along the diagonal, however, correspond to parameter
combinations of $D$ and $H$ which cannot be realized for self-similar
processes.

\bigskip
{\em V. Discussion.} We introduced simple stochastic models which
separate fractal dimension and Hurst coefficient, and allow for any
combination of the two parameters. This is in sharp contrast to
traditional, self-similar models for which fractal dimension and Hurst
coefficient are linearly related. To our knowledge, Figures 1 provides
the first display of fractal images, in which fractal dimension and
Hurst coefficient vary independently. Publicly available code [21]
allows to synthesize images with any pre-specified combination of
fractal dimension and Hurst coefficient. We draw two major
conclusions. The first concerns estimation and measurement. Various
methods have been proposed and applied to estimate fractal dimension
and Hurst coefficient. Popular techniques for estimating or measuring
fractal dimension include box-counting, spectral, and increment-based
methods [1,3,23--25], and estimators for the Hurst coefficient range
from Mandelbrot's R/S analysis to maximum likelihood [1,7,26]. For
estimation of $D$, it is tempting to estimate the Hurst exponent $H$,
and then apply the linear relationship (\ref{eq:DandH}), or vice versa
[27]. We disapprove of any such approach, since the estimator breaks
down easily if the critical assumption of self-similarity is
violated. Secondly, our findings suggest a straightforward test of
self-similarity for time series, profiles, or surfaces [28]: estimate
$D$, a local roughness parameter, and $H$, a long-memory parameter,
and check whether the estimates are statistically compatible with the
linear relationship (\ref{eq:DandH}). A positive answer for a large
number of data sets, across disciplines, will further substantiate the
role of self-similarity within the sciences. Conversely, a negative
answer may reject a self-similar model, but it does not preclude
fractal statistics or long-memory dependence. The Cauchy and modified
Cauchy model provide a striking illustration - and this might be our
key point - that the two notions are independent of each other, and
can be modeled, explained, and synthesized without recourse to
self-similarity.

\smallskip
\subsection*{Acknowledgements} 

Tilmann Gneiting's research has been supported, in part, by the United
States Environmental Protection Agency through agreement CR825173-01-0
to the University of Washington. Nevertheless, it has not been
subjected to the Agency's required peer and policy review and
therefore does not necessarily reflect the views of the Agency and no
official endorsement should be inferred. Martin Schlather has been
supported by the German Federal Ministry of Research and Technology 
(BMFT) through grant PT BEO 51-0339476C.

\bigskip
\natureline

\begin{enumerate}
\item[{[1]}]
B.B.~Mandelbrot, {\em The Fractal Geometry of Nature} (W.~H.~Freeman, 
New York, 1982). 
\item[{[2]}]
B.B.~Mandelbrot, D.E.~Passoja, and A.J.~Paullay, Nature {\bf 308}, 721 
(1984). 
\item[{[3]}]
D.L.~Turcotte, {\em Fractals and Chaos in Geology and Geophysics} (Cambridge 
University Press, Cambridge, 1992).  
\item[{[4]}]
A.~Scotti, C.~Meneveau, and S.G.~Saddoughi, Phys.~Rev.~E {\bf 51}, 5594 
(1995).
\item[{[5]}]
P.~Hall and S.~Davies, Appl.~Phys.~A {\bf 60} (1995).
\item[{[6]}]  
H.E.~Hurst, Trans.~Am.~Soc.~Civil Eng.~{\bf116}, 770 (1951).
\item[{[7]}] 
J.~Beran, {\em Statistics for Long-Memory Processes} (Chapman \und Hall, 
New York, 1994).
\item[{[8]}]
E.~Koscielny-Bunde {\em et al.}, Phys.~Rev.~Lett.~{\bf81}, 729 (1998).    
\item[{[9]}]
H.~Fairfield Smith, J.~Agric.~Sci.~{\bf28}, 1 (1938). 
\item[{[10]}] 
P.~Whittle, Biometrika {\bf 43}, 337 (1956); Biometrika {\bf 49}, 
305 (1962).
\item[{[11]}]
H.A.~Makse, S.~Havlin, M.~Schwartz, and H.E.~Stanley, Phys.~Rev.~E {\bf
53}, 5445 (1996).  
\item[{[12]}]
B.B.~Mandelbrot and J.W.~van Ness, SIAM Rev. {\bf 10}, 422 (1968).
\item[{[13]}]
That is, $< \! Z(x) \! > \; = 0$ and $< \! (Z(x))^2 \! > \; = 1$ for
all $x \in \real^n$, $< \! Z(x)Z(x+h) \! >$ is independent of $x$, and
all marginal distributions are multivariate Gaussian. See A.M. Yaglom,
{\em Correlation Theory of Stationary and Related Random
Functions.~Vol.~I:~Basic Results} (Springer, New York,
1987). Extensions are straightforward, but beyond the scope of this
letter.
\item[{[14]}]
See Chapter 8 of R.J.~Adler, {\em The Geometry of Random Fields}
(Wiley, New York, 1981).
\item[{[15]}]
H.~Wackernagel, {\em Multivariate Geostatistics}, 2nd ed. (Springer, Berlin,
1998); J.-P.~Chil\`es and P.~Delfiner, {\em Geostatistics. Modeling Spatial 
Uncertainty} (Wiley, New York, 1999).  
\item[{[16]}]
A.H.~Romero and J.M.~Sancho, J.~Comput.~Phys.~{\bf156}, 1 (1999).
\item[{[17]}]
D.~Koutsoyiannis, Water Resour.~Res.~{\bf36}, 1519 (2000). 
\item[{[18]}] 
T.~Gneiting, J.~Appl.~Probab.~{\bf37}, 1104 (2000). Arguments along
similar lines show that the given conditions, $\alpha \in (0,2]$ and
$\beta\geq0$, are necessary and sufficient for (\ref{eq:Cauchy}) to be
the correlation function of a stationary random function in $\real^n$.
\item[{[19]}]
O.E.~Barndorff-Nielsen, The.~Probab.~Appl., in press (2001). 
\item[{[20]}]
C.R.~Dietrich, Water Resour.~Res.~{\bf31}, 147 (1995); T.~Gneiting,
Water Resour.~Res.~{\bf32}, 3391 (1996); C.R.~Dietrich, Water
Resour.~Res.~{\bf32}, 3397 (1996).
\item[{[21]}]
M.~Schlather. Contributed package on random field simulation for
R, http://cran.r-project.org/, in preparation.
\item[{[22]}] 
Here we apply the turning bands operator; see, for example, Section 2
of T.~Gneiting, J.~Math. Anal.~Appl.~{\bf236}, 86 (1999). The general
result is that if $c_n(h) = \varphi(|h|)$, $h \in \real^n$, is the
correlation function of a Gaussian random field in $\real^n$ ($n \geq
3$), then there exists a Gaussian random field in $\real^{n-2}$ with
correlation function $c_{n-2}(h) = \varphi(|h|) + |h|/(n-2) \,
\varphi'(|h|)$, $h \in \real^{n-2}$. If $c_n(h)$ is given by
(\ref{eq:Cauchy}), we find that
$$
\textstyle
c_{n-2}(h) = \left( 1 + |h|^\alpha \right)^{-(\beta/\alpha)-1} \left( 1 +
       (1-\frac{\beta}{n-2}) |h|^\alpha \right) \hspace{-0.7mm}, 
       \qquad\quad h \in \real^{n-2},
$$
is a permissible correlation function if $\alpha \in (0,2]$ and $\beta
\geq 0$, with a positive spectral density in $\real^{n-2}$. The modified
Cauchy class (\ref{eq:mod_Cauchy}) corresponds to the special case
when $n=3$. 
\item[{[23]}]
B.~Dubuc {\em et al.}, Phys.~Rev.~A {\bf 39}, 1500 (1989).
\item[{[24]}]
P.~Hall and A.~Wood, Biometrika {\bf 80}, 246 (1993).
\item[{[25]}]
G.~Chan and A.T.A.~Wood, Statist.~Sinica {\bf 10}, 343 (2000).
\item[{[26]}] 
M.J.~Cannon {\em et al.}, Physica A {\bf 241}, 606 (1997). 
\item[{[27]}] 
C.P.~North and D.I.~Halliwell, Math.~Geol.~{\bf26}, 531 (1994). 
\item[{[28]}] 
A wavelet-based test for self-similarity has recently been proposed by
J.-M.~Bardet, J.~Time Ser.~Anal.~{\bf21}, 497 (2000).
\end{enumerate}

\newpage
\section*{Figure legends}

\smallskip
{\bf Figure 1.} Realizations of the Cauchy class with $\alpha = 0.5,
1, 2$ (from left to right) and $\beta = 0.025, 0.2, 0.9$ (from top to
bottom). In each row, the Hurst coefficient, $H = 1-\frac{\beta}{2}$, is
constant, but the fractal dimension, $D = 3-\frac{\alpha}{2}$,
decreases from left to right ($D=2.75,2.5,2$). Accordingly, the images
become smoother. In each column, the fractal dimension is held
constant, but the Hurst parameter decreases from top to bottom
($H=0.9875,0.9,0.55$). Accordingly, persistence and long-range
dependence become less pronounced. The pseudo-random seed is the same
for all nine images, and the length of an edge corresponds to a lag of
16 units in the correlation function (\ref{eq:Cauchy}).

\bigskip
\noindent
{\bf Figure 2.} Realizations of the modified Cauchy class with $\alpha
= 0.65$ (left) and $1.95$ (right), and $\beta = 0.05$ (top,
persistent) and $1.35$ (bottom, antipersistent). Again, the distinct
effects of fractal dimension and Hurst coefficient are evident. The
pseudo-random seed is the same for all four profiles, and the maximal
lag corresponds to 32 units in the correlation function
(\ref{eq:mod_Cauchy}).

\begin{figure}[p]
\centerline{\includegraphics[width=1.97in,angle=270]{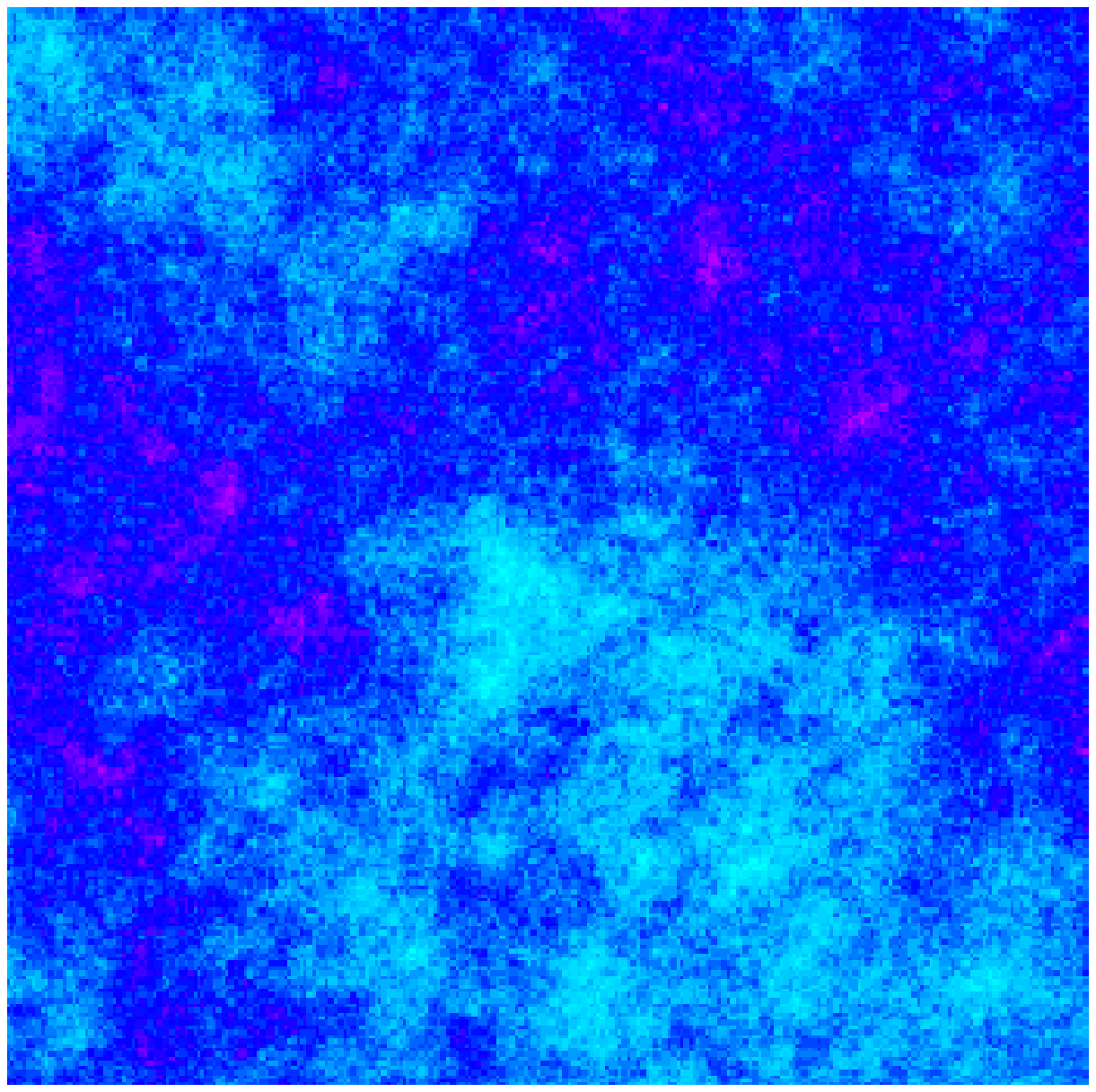}
            \includegraphics[width=1.97in,angle=270]{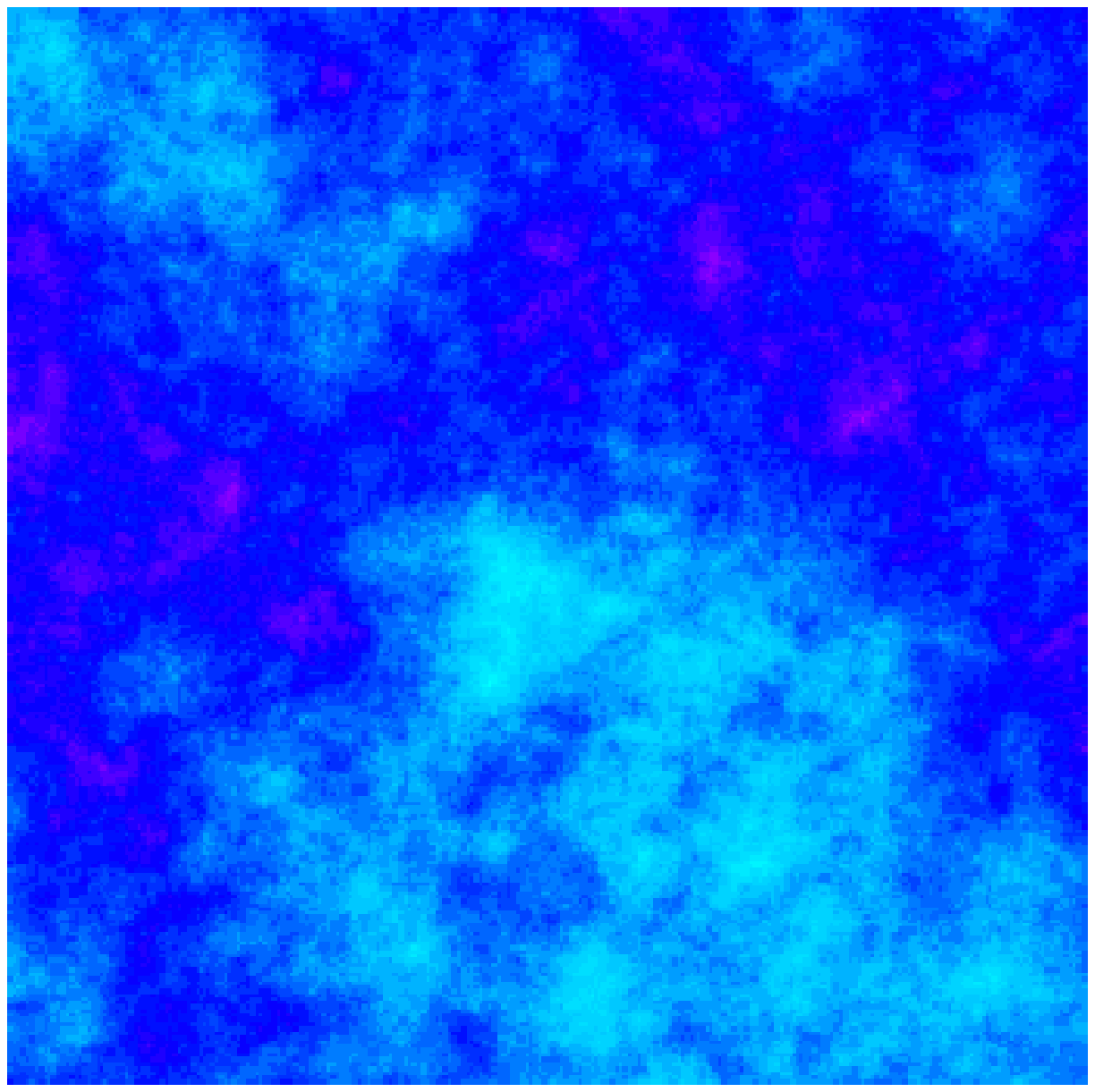}
            \includegraphics[width=1.97in,angle=270]{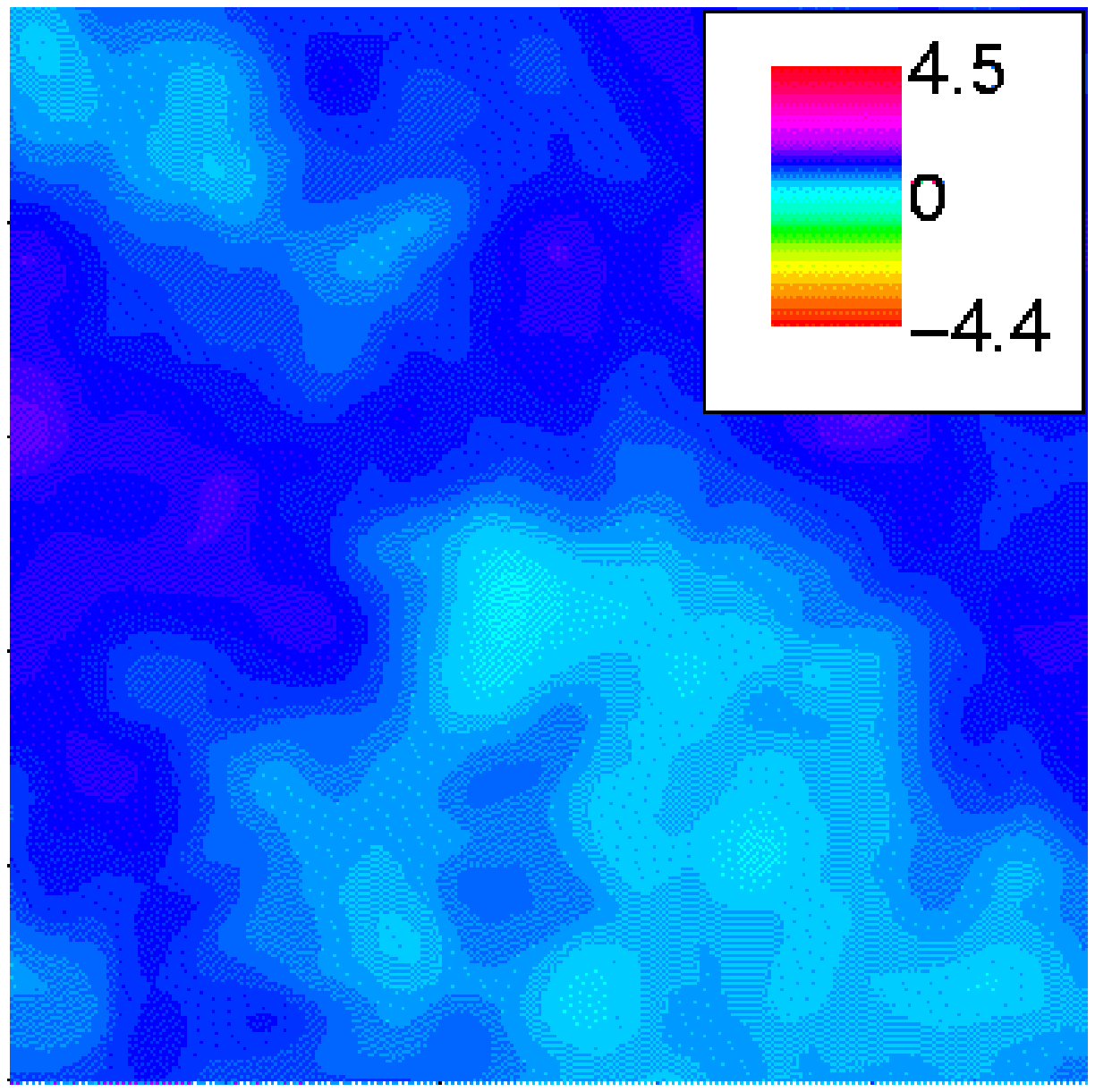}}

\vspace{1mm}
\centerline{\includegraphics[width=1.97in,angle=270]{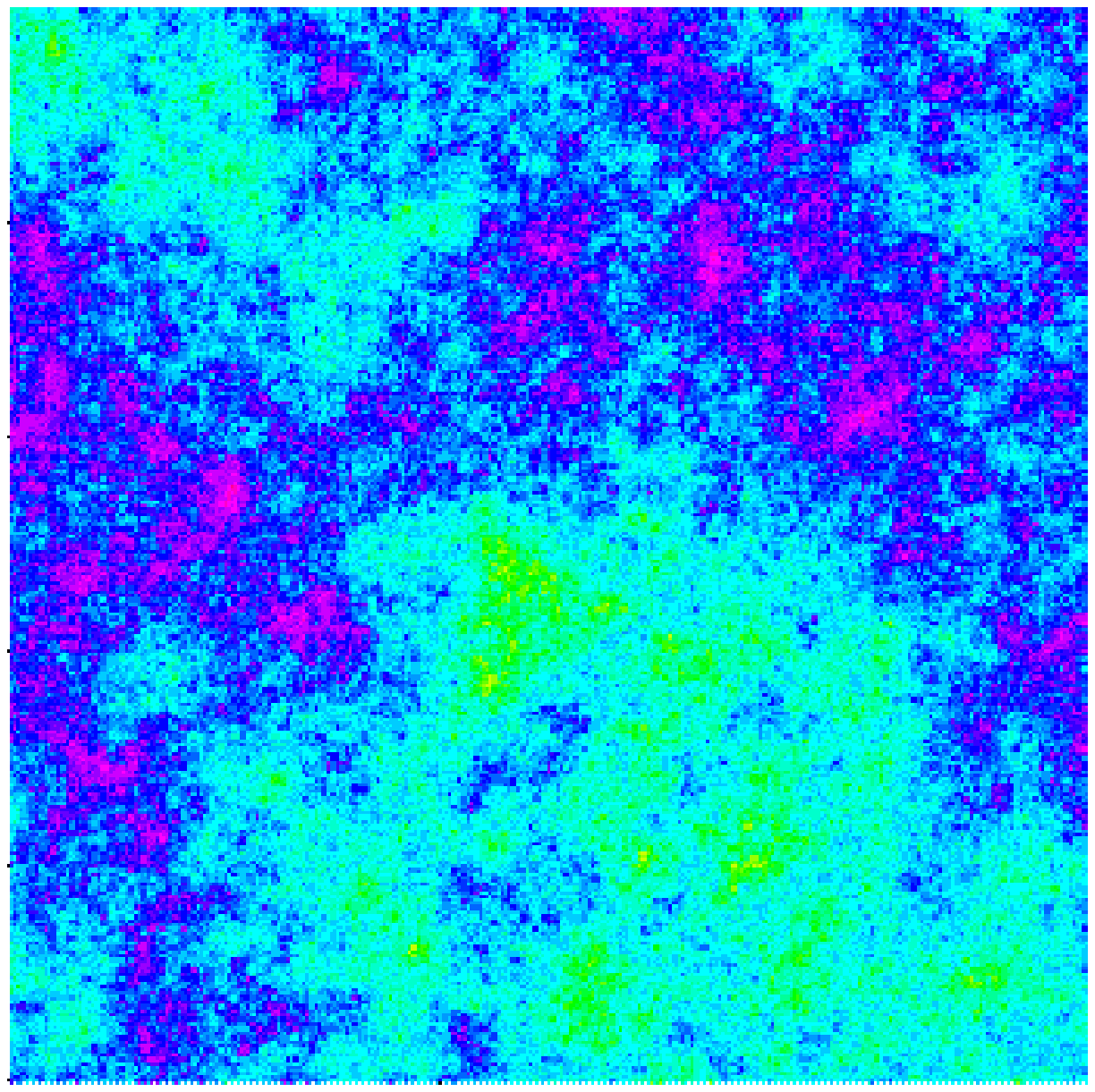}
            \includegraphics[width=1.97in,angle=270]{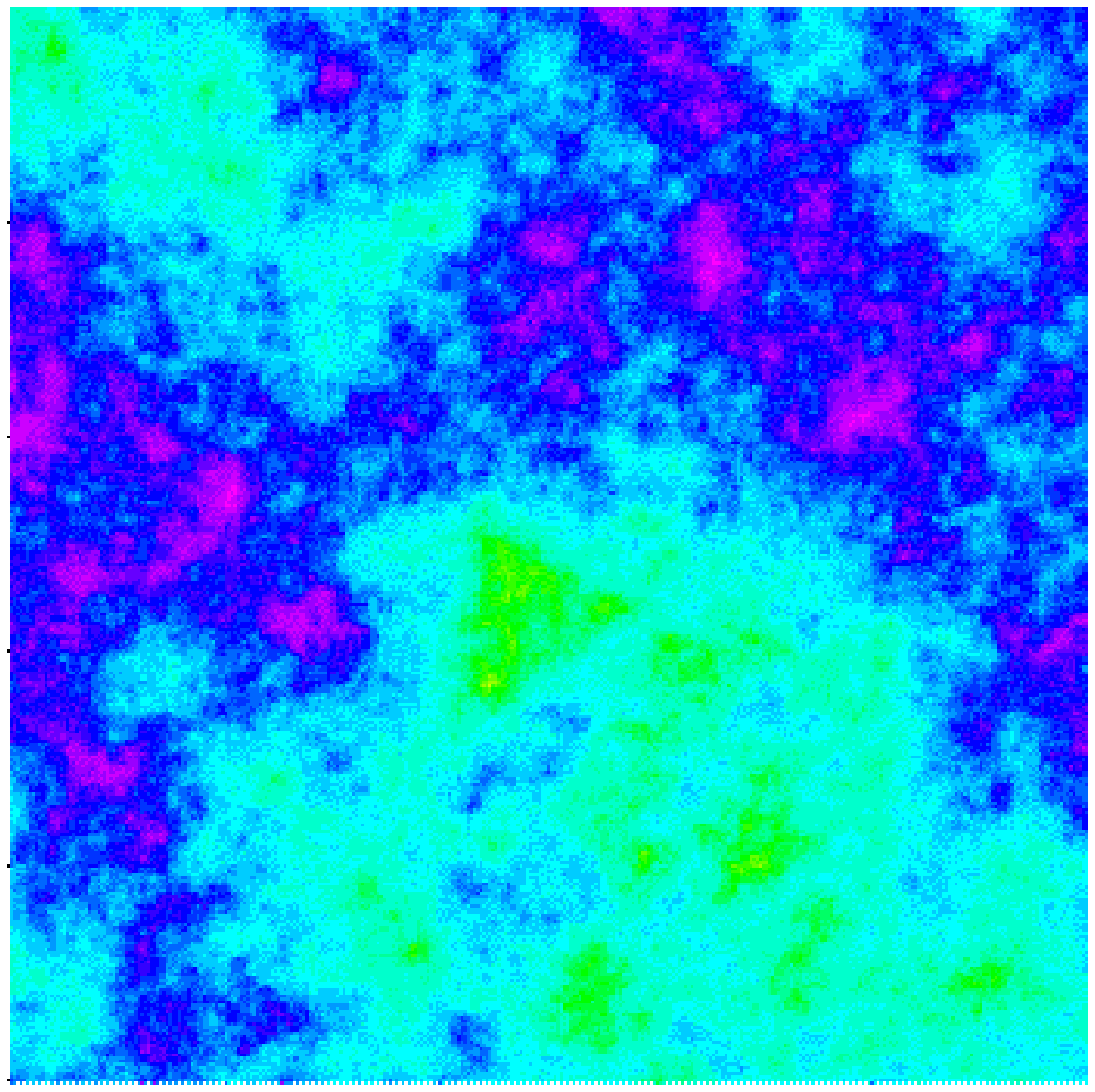}
            \includegraphics[width=1.97in,angle=270]{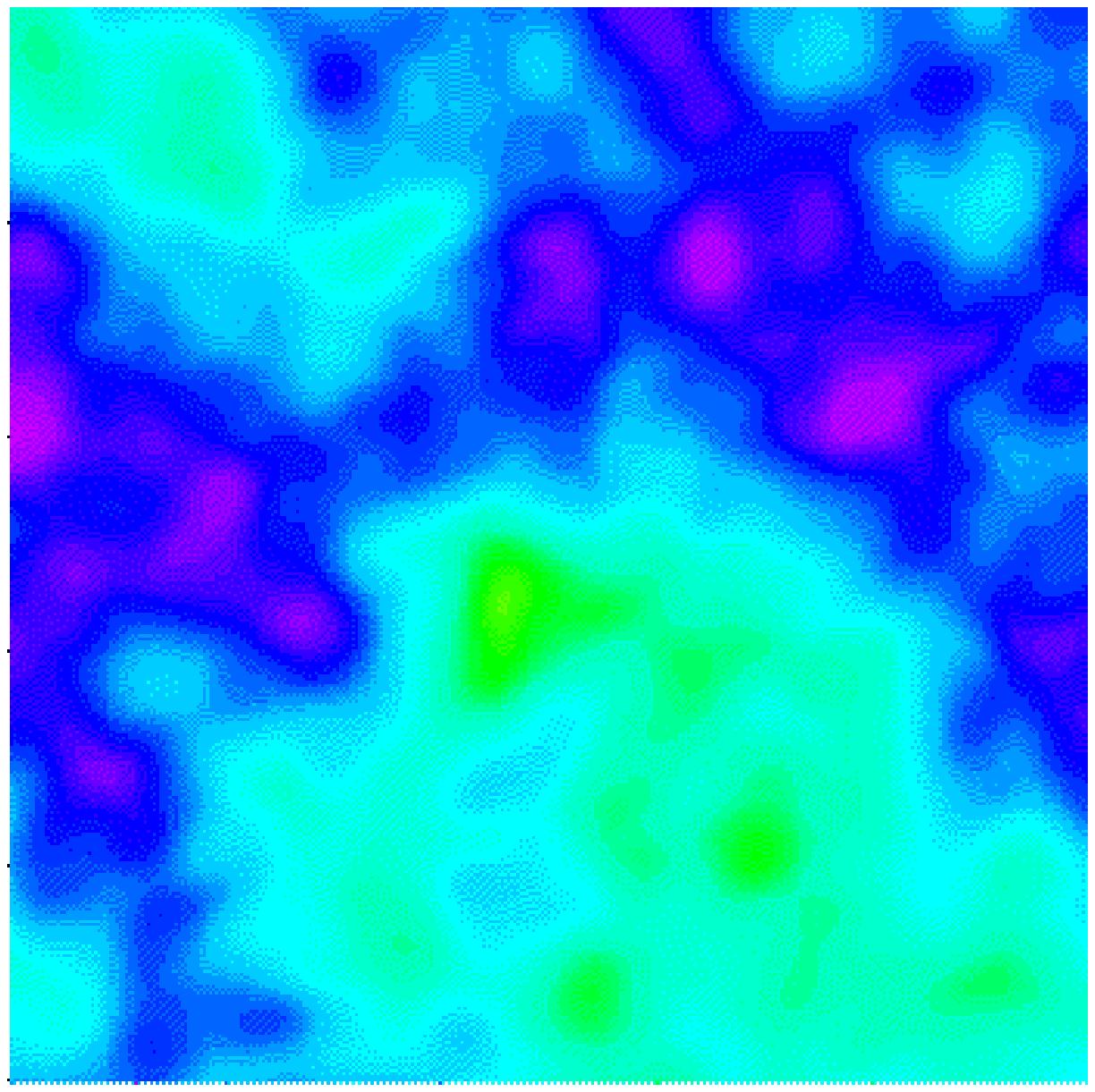}}

\vspace{1mm}
\centerline{\includegraphics[width=1.97in,angle=270]{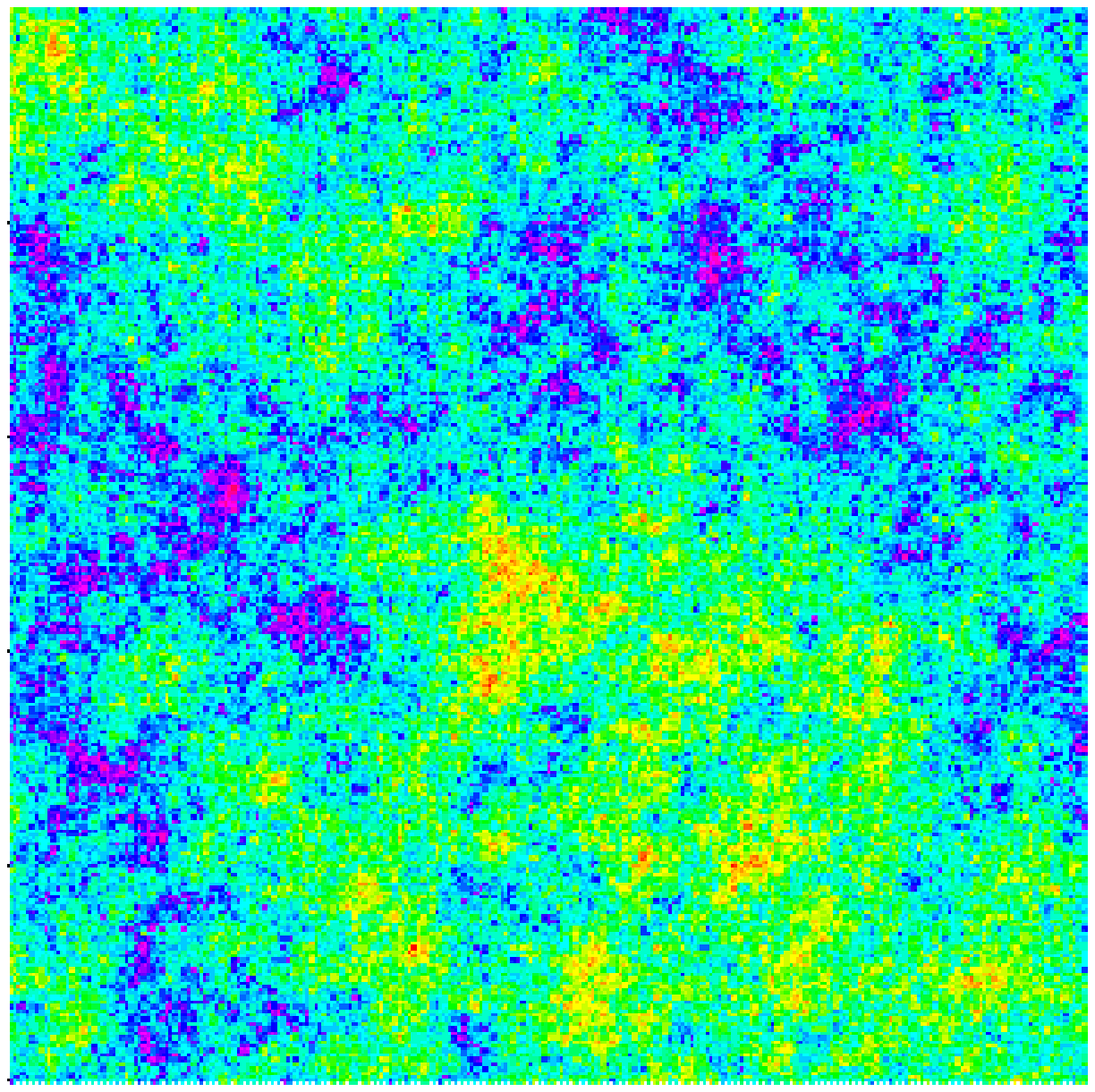}
            \includegraphics[width=1.97in,angle=270]{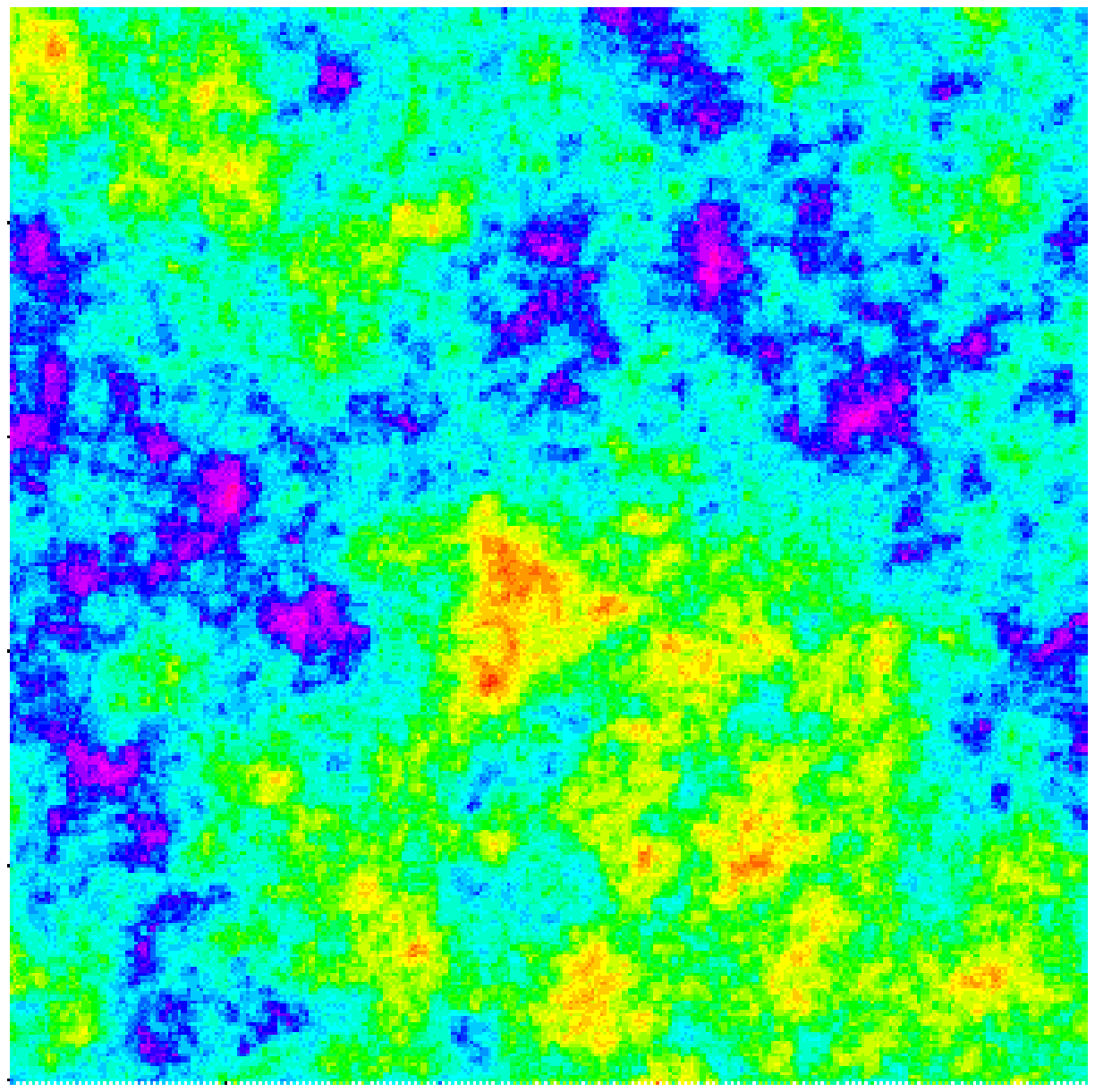}
            \includegraphics[width=1.97in,angle=270]{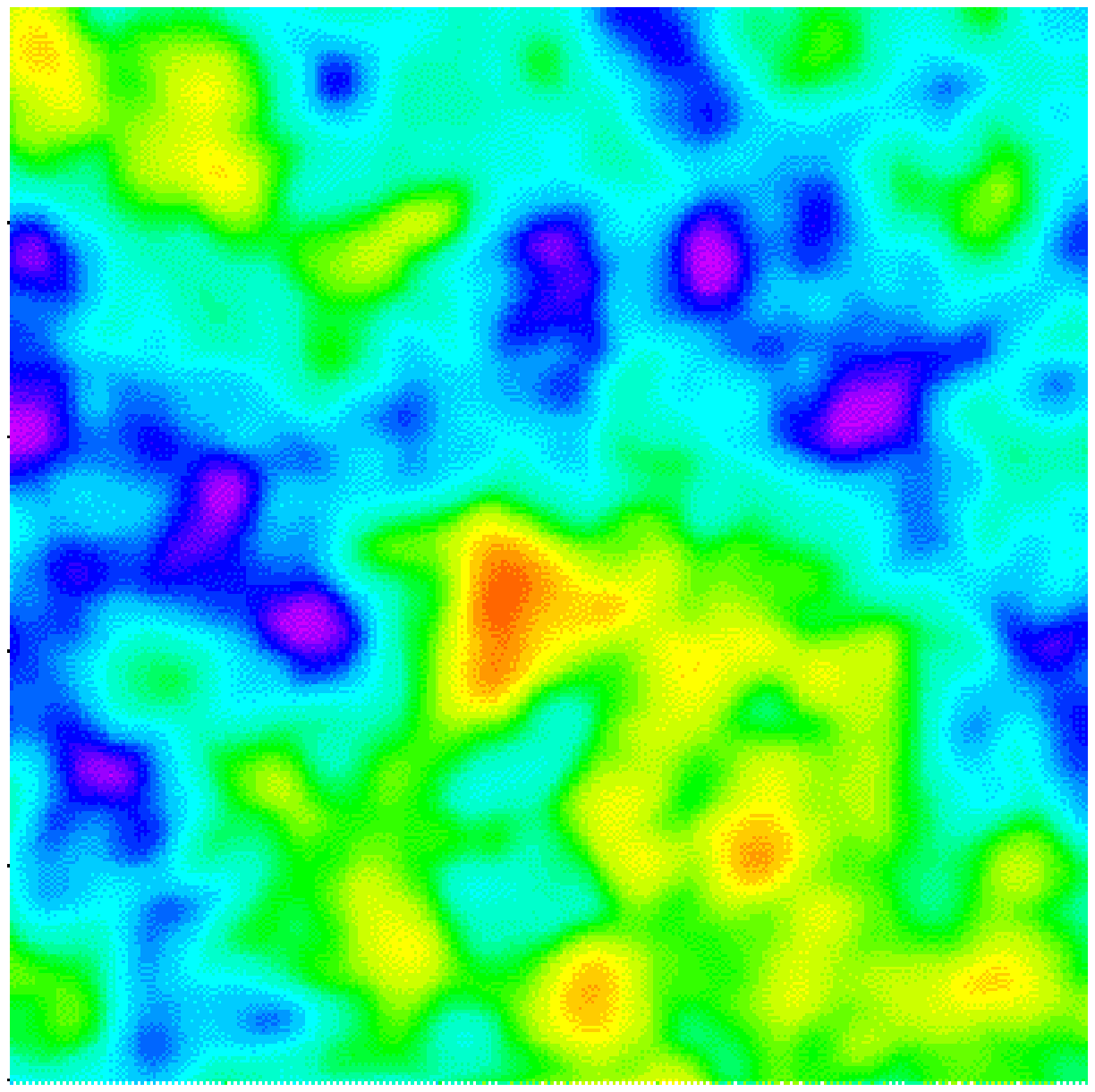}}
\caption{Please use 2 columns.}
\end{figure}

\begin{figure}[p]
\centerline{\scalebox{1.81}[1]{\includegraphics[width=1.3in,angle=270]{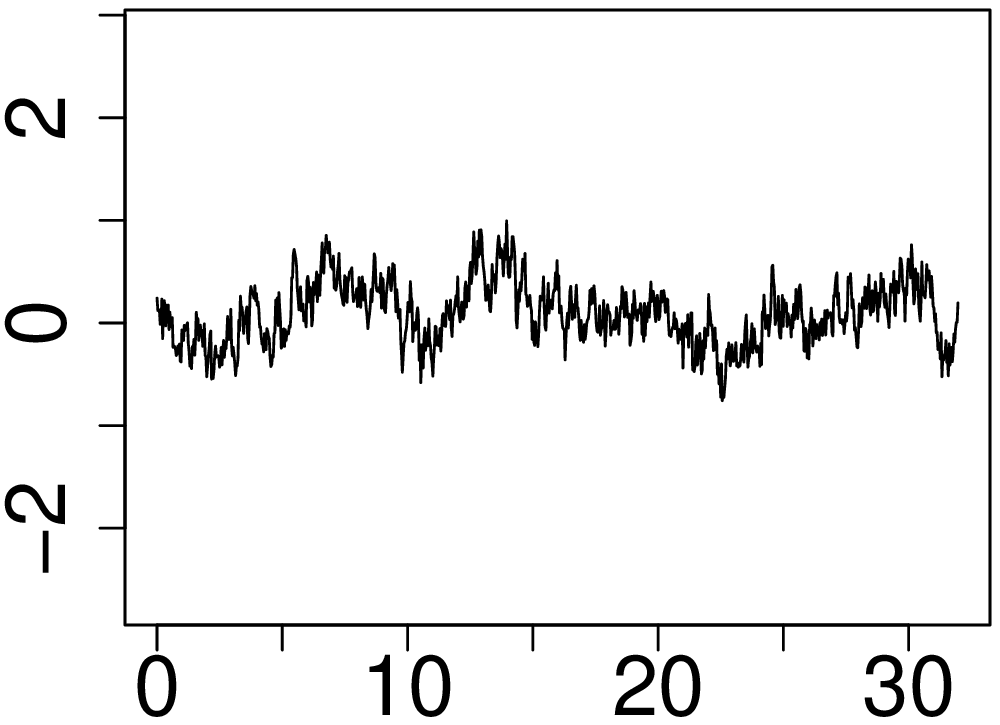}} 
              \hspace{2mm}
            \scalebox{1.81}[1]{\includegraphics[width=1.3in,angle=270]{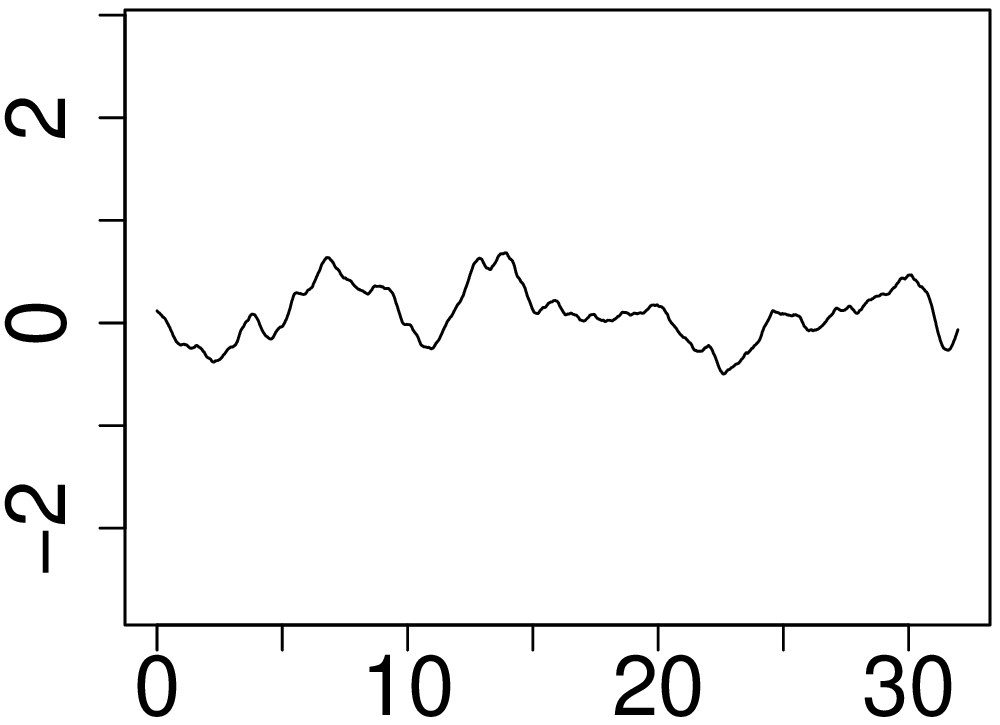}}}

\centerline{\scalebox{1.81}[1]{\includegraphics[width=1.3in,angle=270]{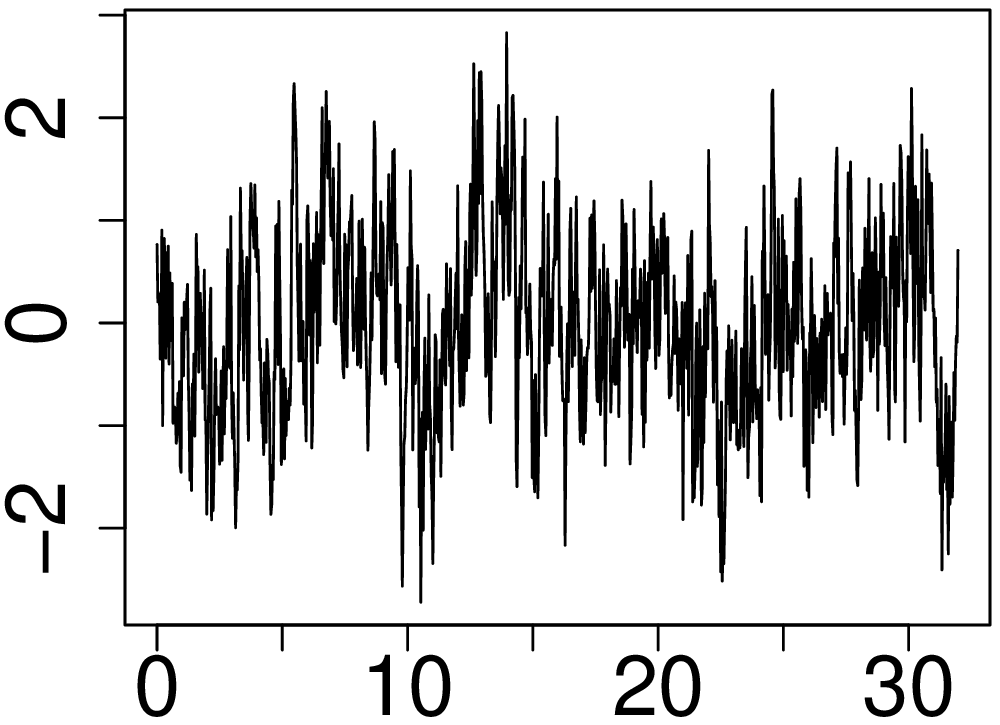}} 
              \hspace{2mm}
            \scalebox{1.81}[1]{\includegraphics[width=1.3in,angle=270]{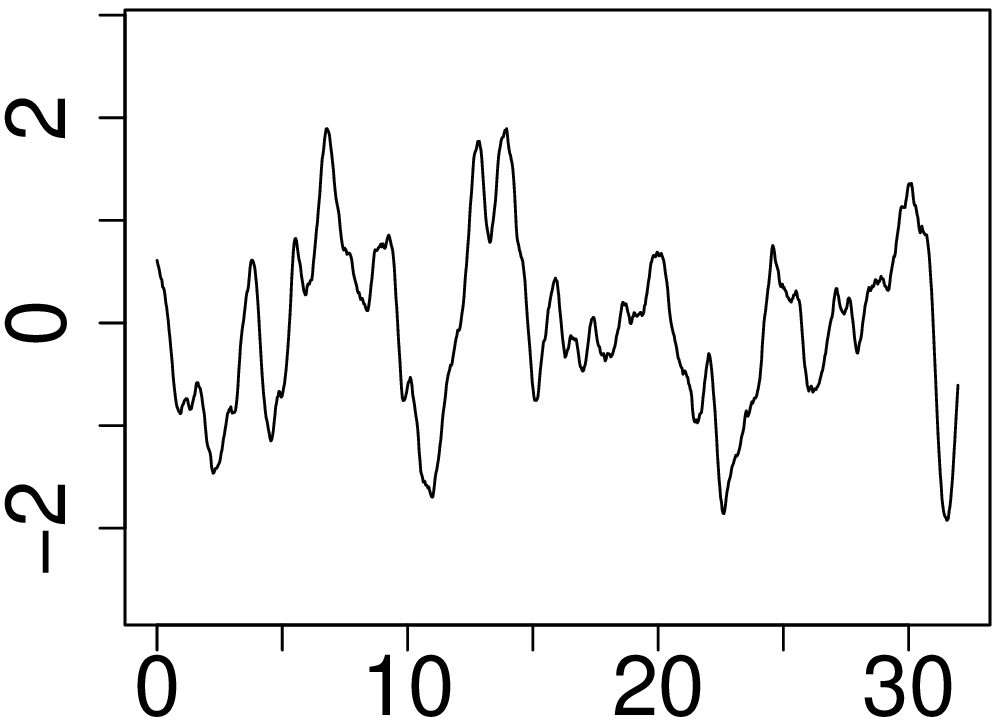}}}
\caption{Please use 1.5 columns.}
\end{figure}

\end{document}